# Thermopower of crown-ether-bridged anthraquinones

Ali K. Ismael[a,b] Iain Grace[a] and Colin J. Lambert[a]

**Abstract** We investigate strategies for increasing the thermopower of crown-ether-bridged anthraquinones. The novel design feature of these molecules is the presence of either (**1**) crown-ether or (**2**) diaza-crown-ether bridges attached to the side of the current-carrying anthraquinone wire. The crown-ether side groups selectively bind alkali- metal cations and when combined with TCNE or TTF dopants, provide a large phase-space for optimising thermoelectric properties. We find that the optimum combination of cations and dopants depends on the temperature range of interest. The thermopowers of both **1** and **2** are negative and at room temperature are optimised by binding with TTF alone, achieving thermpowers of -600 μV/K and -285 μV/K respectively. At much lower temperatures, which are relevant to cascade coolers, we find that for **1**, a combination of TTF and $Na^+$ yields a maximum thermopower of -710 μV/K at 70K, whereas a combination of TTF and $Li^+$ yields a maximum thermopower of -600 μV/K at 90K. For **2**, we find that TTF doping yields a maximum thermopower of -800 μV/K at 90K, whereas at 50K, the largest thermopower (of -600 μV/K) is obtain by a combination TTF and $K^+$ doping. At room temperature, we obtain power factors of 73 μW/m.$K^2$ for **1** (in combination with TTF and $Na^+$) and 90 μW/m.$K^2$ for **2** (with TTF). These are higher or comparable with reported power factors of other organic materials.

## Introduction

A key strategy for improving the thermoelectric properties of inorganic materials has been to take advantage of nanostructuring [1-8]. The thermopower (or Seebeck coefficient) S of a material or nanoscale device is defined as $S = -\Delta V/\Delta T$, where $\Delta V$ is the voltage difference generated between the two ends of a junction when a temperature difference $\Delta T$ is established between them. In addition to the goal of maximising S, there is a world-wide race to develop materials with a high thermoelectric efficiency η. This is expressed in terms of a dimensionless figure of merit $ZT = (P/\kappa)T$, where P is the power factor, T the temperature and κ the thermal conductivity. In terms of ZT, the maximum efficiency of a thermoelectric generator is $\eta_{max} = \eta_C (x-1)/(x+1)$ where $\eta_C$ is the Carnot efficiency and $x = (ZT+1)^{1/2}$, whereas the efficiency at maximum power is $\eta_P = \eta_{CA}(x^2-1)/(x^2+1)$, where $\eta_{CA}$ is the Curzon-Ahlborn upper bound. In both cases, the efficiency is a maximum when ZT tends to infinity.

Nanostructured inorganic materials such as PbSeTe /PbTe-based quantum dot superlattices with ZT of ca. 2 were realised over a decade ago [2]. Since that time the absence of significant improvement in efficiency has led a number of groups to explore thermoelectric properties of single molecules as a first step towards designing higher-performance thermoelectric materials. During the past few years, it has been demonstrated both experimentally and theoretically that at a molecular scale, S can be controlled by varying the chemical composition [9], varying the position of intra-molecular energy levels relative to the work function of metallic electrodes [10], systematically increasing the single-molecule lengths within a family of molecules [11], and by tuning the interaction between two neighbouring molecules [12]. These single-molecule experiments yielded room-temperature values of S ranging in magnitude from ca. 1 to 50 μV/K. They demonstrated an ability to measure single-molecule thermopower, but did not involve a rational design of molecules based on structure-function relationships.

Here we present a detailed study of the electrical conductance and thermopower of molecular junctions with crown-ether moieties, whose structure can be tuned to deliver huge enhancements of thermopower. Crown ethers are cyclic compounds, which consist of a ring of several ether groups. It is known that these structures selectively bind certain cations to form complexes, with the favoured cation determined by the size of the ring. Figure 1 shows examples of our chosen crown ether-bridged molecules, which are composed of an anthraquinone central unit, connected to a crown via either oxygens **1** or nitrogens **2** [13-16].

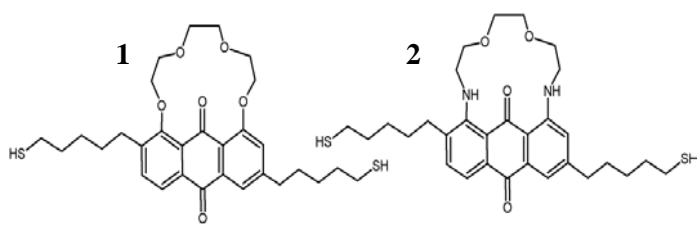

**Figure1**. Molecular wires containing ether bridges. (**left**) AQ15C5 (**right**) AQ diaza-15C5

## Theoretical Method

To calculate the thermopower S and electrical conductance G of these molecules, we use the density functional theory (DFT), code SIESTA [17], combined with the recently-released quantum transport code Gollum [18,19]. The optimum geometries of the isolated molecules were first calculated using density functional theory, with all forces on atoms relaxed to a tolerance of 0.01Volts/Ang. A double zeta plus polarization (DZP) basis set, with norm-conserving pseudo-potentials was employed along with a mesh cutoff of 200 Rydbergs. After optimising their geometries, each molecule was attached to gold leads to form an extended molecule. The leads consisted of seven layers of (111) gold with each layer containing 12 atoms. Gollum uses the DFT-generated Hamiltonian to calculate the zero bias transmission coefficient T(E), from which S and G are given by the following relations:

$$L^n(T) = \int_{-\infty}^{\infty} dE(E-E_F)^n T(E)\left(-\frac{df}{dE}\right) \quad (1)$$

$$S(T) = -\frac{1}{eT}\frac{L^1}{L^0} \quad G(T) = G_0 L^0 \quad (2)$$

where $E_F$ is the Fermi energy, T is the temperature, $f$ is the Fermi function and $G_0$ is the quantum of conductance. To maximise S, our aim is to modify systematically the structures of **1** and **2**, by complexing them with combinations of alkali cations and the donor TTF or the acceptor TCNE. From a mathematical viewpoint, equation (1) tells us that the coefficient $L^1$ and hence S is maximised by creating an asymmetric T(E) in the vicinity of the Fermi energy, because $L^1$ vanishes when T(E) is a symmetric function of $(E - E_F)$. This is achieved when T(E) possesses a sharp maximum located within a few $k_BT$ either above or below the Fermi energy. Chemically, our challenge is to identify optimum combinations of dopants which deliver this feature.

## Results

Figure 2a shows results for the transmission coefficient of the bare molecule **1**, and with a bound potassium cation. For the bare molecule, the LUMO resonance sits approximately 0.12 eV above the DFT-predicted Fermi energy. This is moved to a lower energy upon complexation with the alkali cation. Figure 2b shows the transmission coefficient of **1** before (black) and after complexation with one donor TTF molecule (blue) or one acceptor TCNE molecule (red), and shows that the resonance can be shifted to the right or to left depending on the adsorbate.

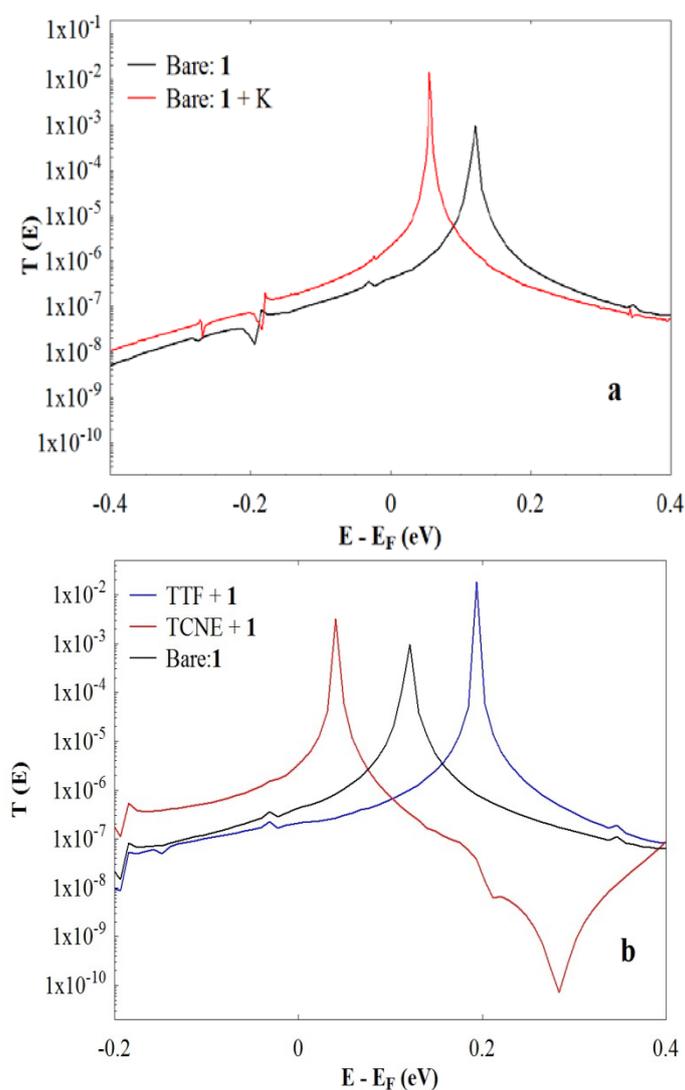

**Figure 2:** The transmission coefficient T(E) for molecule **1** The **upper panel** shows T(E) for the bare molecule (black) and with a bound alkali $K^+$ (red). The **lower panel** shows T(E) for the bare molecule (Black) and with adorbed TCNE (red) or TTF (blue).

To fully explore the range of accessible thermopowers, we have computed the thermopowers of **1** and **2** after binding to different combinations of $K^+$, $Li^+$ or $Na^+$ cations and complexation with TCNE and TTF. To maintain charge neutrality both $PF_6^-$ and $BF_4^-$ were used as counter ions. The predicted thermopowers and electrical conductances were found to be the same for both counter ions. (see SI). In what follows we label each complex as either bare (ie **1** or **2** alone), TTF (ie **1** or **2** with a single adsorbed TTF), TCNE (ie **1** or **2** with a single adsorbed TCNE), TCNE+K (ie **1** or **2** with a single adsorbed TCNE, a single bound $K^+$ and its counter ion $PF_6^-$), TCNE+2K (ie **1** or **2** with a single adsorbed TCNE, two bound $K^+$s and their two counter ions $PF_6^-$), etc.

Figure 3 shows that the thermopower S of all such combinations is negative over a wide range of temperature and that the optimum combination of dopants depends on the temperature range of interest. The thermopowers of both **1** and **2** are negative and at room temperature are optimised by binding with TTF alone, achieving thermopowers of -600 µV/K and -285 µV/K respectively. At much lower temperatures, which are relevant to cascade coolers, we find that for **1**, a combination of TTF and $2Na^+$ yields a maximum thermopower of -710 µV/K at 70K, whereas a combination of TTF and $2Li^+$ yields a maximum thermopower of -600 µV/K at 90K. For **2**, we find that TTF doping yields a maximum thermopower of -800 µV/K at 90K, whereas at 50K, the largest thermopower (of -600 µV/K) is obtain by a combination TTF and $2K^+$ doping.

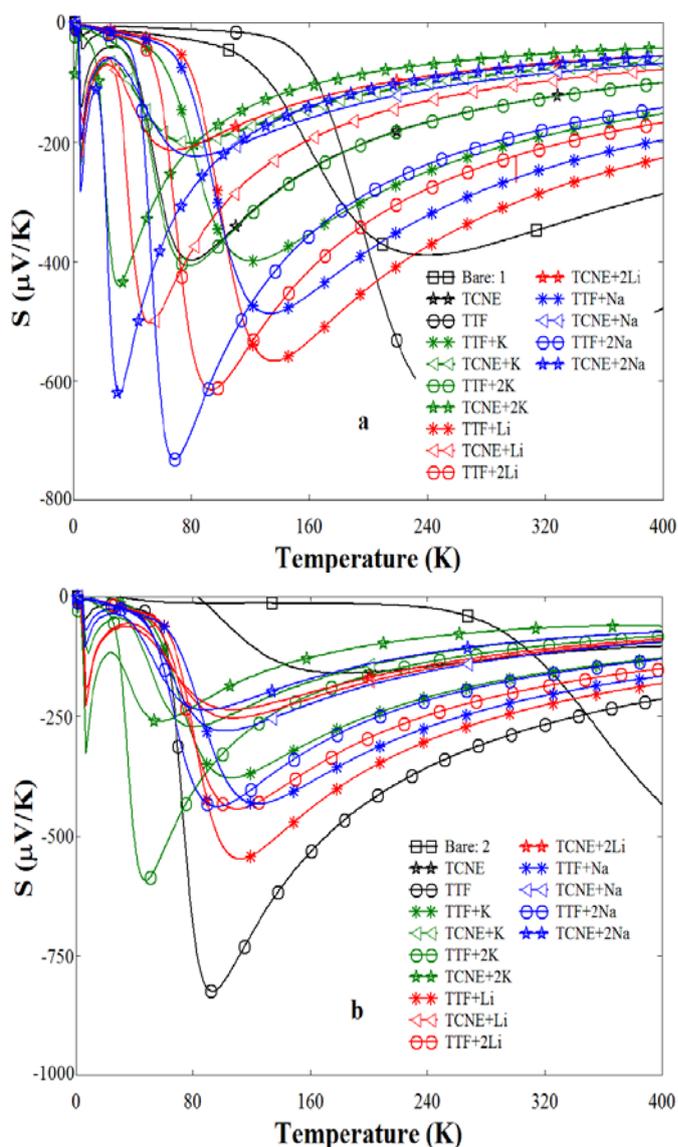

**Figure 3:** Thermopower as function of temperature for complexes of **1** (Fig 3a) and **2** (Fig 3b).

To highlight correlations between the thermopower and electrical conductance, Figure 4 shows plots of thermopower versus electrical conductance at four different temperatures: 250, 300, 350 and 400 K. For all curves apart from bare case (NC) in 3b, the temperature increases from the top (250K) to the bottom (400K). These curves show that for most of the complexes, the thermopower becomes increasingly negative with decreasing temperature apart from the bare **2**. The **1**-TTF has the highest thermopower of -640μV/K at 250K, which is higher than any single-molecule thermopower measured to date.

In practice, the dimensionless figure of merit $ZT = (P/\kappa)T$ is difficult to measure experimentally, because the thermal conductance κ of a single molecule is not directly accessible. However the numerator of ZT (ie the power factor P) is accessible and is defined as $P = S^2\sigma$, where σ is the electrical conductivity. The notion of conductivity is not applicable to transport through single molecules, but to compare with literature values for bulk materials, we define $\sigma = GL/A$, where L and A are equal to the length and the cross-sectional area of the molecule respectively. In what follows, the values $L = 2.45$nm and $A = 7.44$ nm$^2$ are used. From the results of figure 4, the power factors $P = S^2GL/A$ for each of the studied complexes are shown in figure 5.

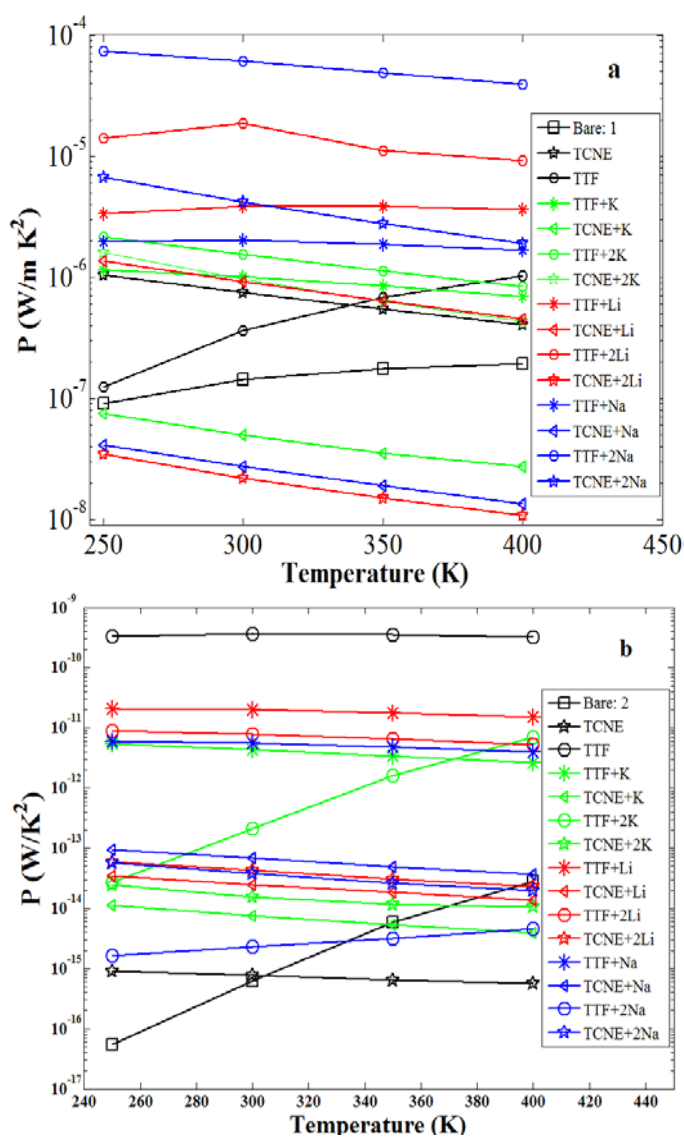

**Figure 5**: Power factor as function of temperature, **top** (**1**) and **bottom** (**2**)

Figure 5a reveals that for **1** complexation with TTF and with TTF + Li yields the highest and next-highest power factors respectively, whereas complexation with TCNE has the lowest. Figure 5b shows that for **2**, complexation with TTF + 2Na and TTF + 2Li yield the highest and next-highest power factors respectively, while complexation with TCNE + 2Li yields the lowest. In all cases, these reflect the positions of transmission resonances relative to the Fermi energy.

## Conclusions

The thermoelectric properties of a series of molecular wires have been calculated in the presence of one or two alkali cations and in the presence of acceptor (TCNE) and donor (TTF) molecules. Due to charge transfer between the host and adsorbates, complexation produces a shift in the LUMO transmission resonances, resulting in orders of magnitude variation in both the thermopower and power factor. In the case of TTF the resonance shifts away from the Fermi energy, leading to a decrease in the value of S, while for TCNE, the resonance shifts closer to the Fermi energy, causing the thermopower to increase. At room temperature we find that **1** doped with TTF possesses the largest calculated thermopower of -640 µV/K, which is higher than any single-molecule thermopower measured to date. At room temperature, we obtain power factors of 73 µW/m.K$^2$ for **1**+TTF+2Na and 90 µW/m.K$^2$ for **2**+TTF. These compare favourably with power factors of other organic materials, whose reported values range from 0.016 µW/m.K$^2$ and 0.045 µW/m.K$^2$ for Polyaniline and Polypyrole respectively [20], to 12 µW/m.K$^2$ for PEDOT:PSS[21] and (12 µW/m.K$^2$ for $C_{60}$/$Cs_2Co_3$ Dph-BDT [22].

## Acknowledgements

This work was supported by the Swiss National Science Foundation (No. 200021-147143) as well as by the European Commission (EC) FP7 ITN "MOLESCO" project no. 606728 and UK EPSRC, (grant nos. EP/K001507/1, EP/J014753/1, EP/H035818/1), and the Iraqi Ministry of Higher Education, Tikrit University (SL-20). A. K. I. acknowledges financial support from Tikrit University.